\documentclass[twocolumn,amssymb, amsmath, aps, superscriptaddress, showpacs, footinbib,  prb]{revtex4}
\usepackage{graphicx}

\newcommand{\reg}{\text{reg}}

\begin{document}

\title{Exploring the spin-$\frac 12$ frustrated square lattice model\\ with high-field magnetization measurements}
\author{Alexander A. Tsirlin}
\email{altsirlin@gmail.com}
\affiliation{Max Planck Institute for Chemical Physics of Solids, N\"{o}thnitzer
Str. 40, 01187 Dresden, Germany}
\affiliation{Department of Chemistry, Moscow State University, 119991 Moscow, Russia}
\author{Burkhard Schmidt}
\affiliation{Max Planck Institute for Chemical Physics of Solids, N\"{o}thnitzer
Str. 40, 01187 Dresden, Germany}
\author{Yurii Skourski}
\affiliation{Dresden High Magnetic Field Laboratory, Forschungszentrum Dresden-Rossendorf, 01314 Dresden, Germany}
\author{Ramesh Nath}
\altaffiliation[Present address: ]{Ames Laboratory and Department of Physics and Astronomy, Iowa State University, Ames, Iowa 50011 USA}
\affiliation{Max Planck Institute for Chemical Physics of Solids, N\"{o}thnitzer
Str. 40, 01187 Dresden, Germany}
\author{Christoph Geibel}
\author{Helge Rosner}
\email{Helge.Rosner@cpfs.mpg.de}
\affiliation{Max Planck Institute for Chemical Physics of Solids, N\"{o}thnitzer
Str. 40, 01187 Dresden, Germany}

\begin{abstract}
We report on high-field magnetization measurements for a number of layered vanadium phosphates that were recently recognized as spin-1/2 frustrated square lattice compounds with ferromagnetic nearest-neighbor couplings ($J_1$) and antiferromagnetic next-nearest-neighbor couplings ($J_2$). The saturation fields of the materials lie in the range from 4 to 24~T and show excellent agreement with the previous estimates of the exchange couplings deduced from low-field thermodynamic measurements. The consistency of the high-field data with the regular frustrated square lattice model provides experimental evidence for a weak impact of spatial anisotropy on the nearest-neighbor couplings in layered vanadium phosphates. The variation of the $J_2/J_1$ ratio within the compound family facilitates the experimental access to the evolution of the magnetization curve upon the change of the frustration magnitude. Our results support the recent theoretical prediction by Thalmeier \textit{et~al.} [Phys. Rev. B, \textbf{77}, 104441 (2008)] and give evidence for the enhanced bending of the magnetization curves due to the increasing frustration of the underlying spin system.
\end{abstract}

\maketitle

Quantum phenomena and exotic ground states in low-dimensional magnets receive broad attention from both experimental and theoretical sides. The application of an external magnetic field provides a natural and experimentally feasible way to alter the properties of such systems. Recent studies showed that numerous low-dimensional magnets undergo field-induced phase transitions that can be described with bosonic models.\cite{giamarchi} A case of particular interest are the frustrated spin systems, because frustration often leads to the localization of magnetic excitations on the lattice and can cause the formation of unusual ordered phases, evidenced by plateau features on the magnetization curves. 

\begin{figure}
\includegraphics{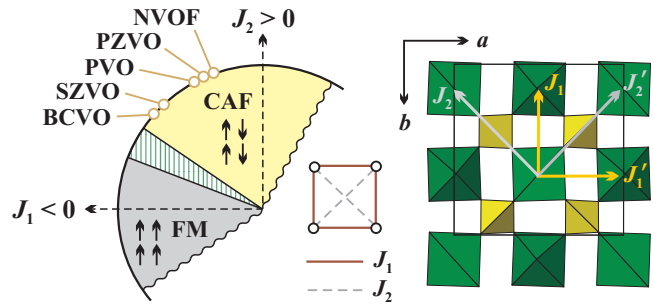}
\caption{\label{diagram}
Left panel: part of the ground state phase diagram for the FSL model;\cite{shannon2004} solid filling shows the regions of the FM and CAF ordering, while hatched filling denotes the critical region around $\alpha=-0.5$. Right panel: the [VOPO$_4$] layer in the structures of layered vanadium phosphates. In the left panel, the model compounds under discussion are placed into the phase diagram according to Table~\ref{table}.
}
\end{figure}

The frustrated square lattice (FSL) is one of the most simple low-dimensional frustrated spin models. This model assumes the competing interactions along the sides and the diagonals of the square, $J_1$ and $J_2$, respectively (see Fig.~\ref{diagram}). The variation of the frustration ratio $\alpha=J_2/J_1$ enables to tune the system from a non-frustrated to a strongly frustrated regime. For the case of spin-$\frac12$, theoretical studies suggest three ordered phases and two critical spin-liquid regions in the ground state phase diagram.\cite{misguich,shannon2004} Model simulations propose the formation of a magnetization plateau at half-saturation\cite{zhitomirsky2000} (and, possibly, another plateau at one-third of the saturated magnetization\cite{chang2002}) for the region of $J_2/J_1\simeq 0.5$. Later theoretical studies focused on the magnetocaloric effect, the saturation fields, and the shape of the magnetization curves for the whole phase diagram.\cite{burkhard2007,thalmeier2008} In particular, frustration-driven quantum fluctuations were shown to enhance the bending of the magnetization curves as compared to the linear curve for the classical system.\cite{thalmeier2008} Despite the broad theoretical interest, systematic experimental high-field studies of FSL compounds have not been reported.

Recently, a number of spin-$\frac12$ FSL model compounds were proposed.\cite{melzi2000,helge2002,carretta2002,kaul2004,tsirlin2008} Most of these materials reveal relatively weak exchange couplings, thus efficient high-field experiments in pulsed or even static fields are feasible. In this Brief Report, we will focus on the family of layered vanadium phosphates that lie in the region of columnar antiferromagnetic (CAF) ordering in the ferromagnetic (FM) $J_1$ -- antiferromagnetic (AFM) $J_2$ part of the general phase diagram (see Fig.~\ref{diagram}). Below, we will discuss five compounds of this family. The chemical compositions of these compounds are listed in Table~\ref{table} along with their abbreviations. 

The low-field properties of PVO and SZVO have been reported in Refs.~\onlinecite{kaul2004} and~\onlinecite{enrique}. The compounds NVOF and PZVO were recognized as the FSL materials quite recently, and the detailed study of their low-field behavior will be published elsewhere. The investigation of BCVO has been reported in Ref.~\onlinecite{nath2008} along with the full magnetization curve that will be used here for the sake of comparison. The high-field data for the other four compounds are original and have not been previously reported.

\begin{table*}
\caption{\label{table}
Averaged NN ($\bar J_1$) and NNN ($\bar J_2$) exchange couplings in layered vanadium phosphates according to the low-field thermodynamic measurements, the frustration ratio $\alpha$, thermodynamic energy scale $J_c=\sqrt{\bar J_1^2+\bar J_2^2}$, the experimental saturation field ($\mu_0H_s$), and the calculated saturation field for the regular FSL model ($\mu_0H_s^{\reg}$), as determined from Eq.~\eqref{saturation} using the $\bar J_1$ and $\bar J_2$ values.
}
\begin{ruledtabular}
\begin{tabular}{c@{}cccccccc}
  Index & & Compounds & $\bar J_1$ (K) & $\bar J_2$ (K) & $\alpha=\bar J_2/\bar J_1$ & $J_c$ (K) & $\mu_0H_s$ (T) & $\mu_0H_s^{\reg}$ (T) \\
  NVOF & & Na$_{1.5}$VOPO$_4$F$_{0.5}$ & $-3.2$ & $6.3$ & $-2.0$ & 7.1 & 15.4 & 14.3 \\
  PZVO & & PbZnVO(PO$_4)_2$ & $-5.2$ & $10.0$ & $-1.9$ & 11.3 & 23.4 & 22.6 \\  
  PVO & & Pb$_2$VO(PO$_4)_2$ & $-5.1$ & $9.4$ & $-1.8$ & 10.7 & 20.9 & 20.9 \\
  SZVO & & SrZnVO(PO$_4)_2$ & $-8.3$ & 8.9 & $-1.1$ & 12.2 & 14.2 & 14.5 \\
  BCVO & (Ref.~\onlinecite{nath2008}) & BaCdVO(PO$_4)_2$ & $-3.6$ & 3.2 & $-0.9$ & 4.8 & 4.2 & 4.3 \\
\end{tabular}
\end{ruledtabular}
\end{table*}

Layered vanadium phosphates have a common structural feature, the [VOPO$_4$] magnetic layers with V$^{+4}$O$_5$ square pyramids and superexchange pathways for the nearest-neighbor (NN) and next-nearest-neighbor (NNN) interactions (see Fig.~\ref{diagram}). The low crystal symmetry usually leads to the non-equivalent couplings along the sides ($J_1,J_1'$) and the diagonals ($J_2,J_2'$) of the square, thus a rigorous spin model including first and second neighbors should comprise four independent parameters. However, our recent computational study\cite{tsirlin2009} demonstrated that this effect of the \emph{spatial anisotropy} is usually weak. In particular, thermodynamic properties of the layered vanadium phosphates can be accurately described with the regular FSL model and do not allow to estimate the magnitude of the spatial anisotropy. In contrast, the field dependence of the magnetization should be altered by the spatial anisotropy of nearest-neighbor couplings. This problem will be further discussed below. 

The low-field thermodynamic data yield averaged nearest-neighbor and next-nearest-neighbor couplings [$\bar J_1=(J_1+J_1')/2$ and $\bar J_2=(J_2+J_2')/2$, respectively] for the distorted square lattice.\cite{tsirlin2009} The resulting estimates of $\bar J_1$ and $\bar J_2$ are listed in Table~\ref{table}. These numbers can be used to calculate the effective frustration ratio $\alpha=\bar J_2/\bar J_1$ and to place the compounds into the general phase diagram (Fig.~\ref{diagram}).

Polycrystalline samples of PVO and SZVO were prepared according to Ref.~\onlinecite{enrique}. The powder samples of NVOF and PZVO were obtained by firing the mixtures of NaPO$_3$, NaF, and VO$_2$ (NVOF) or PbZnP$_2$O$_7$, V$_2$O$_3$, and V$_2$O$_5$ (PZVO) in evacuated silica tubes at 700~$^{\circ}$C. The details of the preparation procedure will be reported elsewhere. The phase purity was checked using X-ray diffraction. All the samples except for PZVO were single-phase. The sample of PZVO contained $2-3$\% of the diamagnetic PbZnP$_2$O$_7$ impurity that did not influence on the shape of the magnetization curve. 

Magnetization measurements were performed in the Dresden High Magnetic Field Laboratory using a pulsed magnet powered by an 1.44 MJ capacitor bank. With an inner bore of 20 mm, the magnet yielded fields up to 60 Tesla with a rise time of 7 ms and the total pulse duration of about 20 ms. Magnetic moment of the sample was obtained by integration of the voltage induced in a compensated pick-up coil system surrounding the sample. Each measurement was performed twice, with and without the sample, in order to remove the background signal contribution. All the data were collected at a temperature of 1.4~K. The curves measured on increasing and decreasing field coincided, thus indicating the lack of any irreversible effects upon magnetization and demagnetization of the samples.

\begin{figure}
\includegraphics{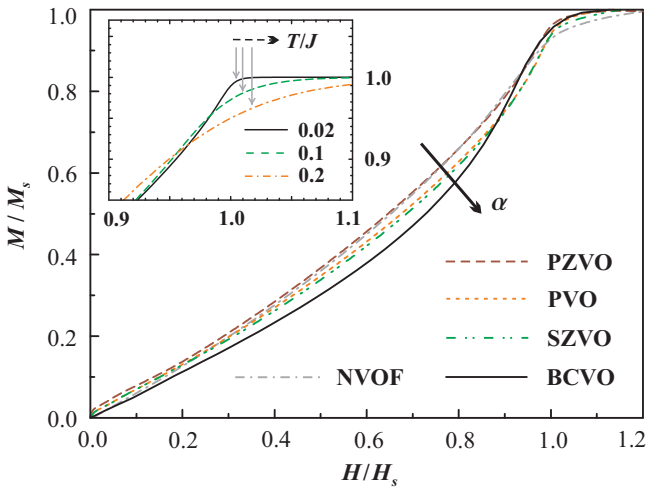}
\caption{\label{curves}
Experimental magnetization curves normalized to the saturation field ($H_s$) and to the saturated magnetization ($M_s$), the arrow shows the evolution of the curves upon the increase of the frustration ratio $\alpha=\bar J_2/\bar J_1$. The inset shows QMC simulations for the unfrustrated square lattice with the exchange coupling $J$. Solid, dashed, and dash-dotted curves refer to the temperatures ($T/J$) of 0.02, 0.1, and 0.2. Solid arrows denote the points of the maximum curvature, while the dashed arrow shows the shift of these points with the increase of temperature.
}
\end{figure}

All the four compounds show monotonous increase of the magnetization ($M$) upon the application of the external magnetic field ($H$). The saturation is evidenced by a sharp bend of the curve (see Fig.~\ref{curves}). Above this anomaly, the magnetization reaches the saturated value ($M_s$) and remains 
field-independent%
.

The saturation field ($H_s$) is usually determined by taking derivatives of finite-temperature magnetization curves. The $H_s$ value is assigned to the inflection point of the first derivative or to the minimum of the second derivative (both refer to the point of the maximum curvature for the $M$ vs. $H$ dependence). The saturation fields obtained in this way are listed in Table~\ref{table}. The uncertainty of the resulting numbers depends on two factors: (i) the accuracy of the experimental data; (ii) the temperature of the measurement. Experiments in pulsed fields yield a very high field resolution; therefore, the calculation of the derivatives is quite accurate, and the uncertainty from (i) does not exceed 0.1~T. The effect of (ii) is a smoothening of the saturation anomaly. At finite temperatures, the maximum magnetization is reached above $H_s$, because an additional energy (and, hence, higher field) is required to suppress thermal fluctuations. 

To illustrate the finite-temperature effect and to estimate the resulting error in the $H_s$ values, we simulated magnetization curves for a representative two-dimensional spin model, the unfrustrated square lattice with the single parameter $J=J_1$ ($J_2=0$). The simulations were performed using the ALPS code\cite{alps} and the quantum Monte-Carlo (QMC) directed loop algorithm in the stochastic series expansion representation.\cite{dirloop} The QMC magnetization curves for the unfrustrated square lattice are shown in the inset of Fig.~\ref{curves}. One can see that thermal fluctuations tend to smoothen the curve, and the saturated value $M_s$ is reached well above $H_s$. Still, the points of the maximum curvature remain close to the true saturation field for $T=0$, although the experimental $H_s$ numbers at finite $T$ should be slightly overestimated. For example, the error amounts to about 1\% of $H_s$ at $T=0.1J$ and to about 2\% of $H_s$ at $T=0.2J$. For the frustrated square lattice, the energy scale is defined by the value of $J_c=\sqrt{\bar J_1^2+\bar J_2^2}$. The experiments for PZVO, PVO, and SZVO are performed at $T\simeq 0.1J_c$, thus the uncertainty of $H_s$ amounts to $0.15-0.20$~T. In case of NVOF, the uncertainty may be slightly higher due to the lower $J_c$. The magnetization curve of BCVO has been collected at 0.5~K\cite{nath2008} that also corresponds to $T\simeq 0.1J_c$. Overall, the total error from (i) and (ii) should not exceed 0.3~T for all the compounds. 

After establishing the uncertainty of the experimental $H_s$ values, one can compare them to the low-field estimates of the exchange couplings. All the compounds under investigation lie in the region of CAF ordering (see Fig.~\ref{diagram}), hence the saturation field of the regular FSL amounts to\cite{burkhard2007} 
\begin{equation}
  \mu_0H_s=4S(\bar J_1+2\bar J_2)k_B/(g\mu_B),
\label{saturation}\end{equation}  
where $S=\frac12$ is the spin value, $k_B$ is the Boltzmann constant, $\mu_B$ is Bohr magneton, and $g$ is the $g$-factor. In the following, we assume $g=1.95$ as a representative value for V$^{+4}$-containing compounds (see, e.g., Ref.~\onlinecite{sr2v3o9-ESR}).

The experimental $H_s$ values and the low-field estimates are compared in Table~\ref{table}. For most of the compounds, we find perfect agreement with the previously obtained $\bar J_1$ and $\bar J_2$ values. In case of NVOF and PZVO, the exchange integrals slightly underestimate $H_s$, and this effect will be further discussed below. A comparison of the saturation fields manifests the role of the frustration in layered vanadium phosphates. The saturation field of SZVO is well below that for PZVO and PVO, although SZVO shows the largest $J_c$ among the whole family. This conclusion is consistent with the previous phenomenological evidence of the frustration, e.g., the suppression of the specific heat maximum in SZVO as compared to PVO.\cite{enrique}

The saturation fields can also be used to estimate the spatial anisotropy of nearest-neighbor couplings in layered vanadium phosphates. Above, we have mentioned that the low crystal symmetry leads to inequivalent exchange couplings along the sides and the diagonals of the square (Fig.~\ref{diagram}). In case of the CAF ordering, the difference between $J_1$ and $J_1'$ should alter the saturation field. If $J_1$ and $J_1'$ are FM, and $|J_1'|<|J_1|$, the columnar ordering is favored by $J_1, J_2$, and $J_2'$, while disfavored by $J_1'$. Thus, the reduction of $|J_1'|$ should stabilize the columnar ordering with respect to the ferromagnetic state, thus leading to the increase of $H_s$. Using the classical Heisenberg model, one easily finds that the increase of the true saturation field ($H_s$) as compared to the estimate for the regular lattice ($H_s^{\reg}$) is proportional to $(J_1'-J_1)$: $\mu_0H_s-\mu_0H_s^{\reg}=2S(J_1'-J_1)k_B/(g\mu_B)$. For PVO, SZVO, and BCVO, the $H_s$ and $H_s^{\reg}$ values nearly match. Yet PZVO shows the anisotropy $J_1'-J_1\simeq 0.8$~K. In case of NVOF, the anisotropy is even more pronounced: $J_1'-J_1\simeq 1.4$~K. The origin of the spatial anisotropy in these compounds is still to be clarified.

Now, we turn to the shapes of the experimental magnetization curves. Fig.~\ref{curves} presents a comparative plot with all the curves normalized to the saturation field and to the saturated magnetization. To determine $M_s$, we used $M$ values well above $H_s$%
. One can observe a clear difference between the curves. As the frustration is increased, the bending gets more pronounced. The strongest bending is found for BCVO with the largest frustration ratio $\alpha$ and the strongest frustration. This result confirms the recent theoretical prediction on the quantum effects in the high-field magnetization of the FSL model.\cite{thalmeier2008} Qualitatively, the effect can be understood as the competition of quantum fluctuations and the external field. Frustration induces strong quantum fluctuations that reduce magnetization values at low fields and cause the positive curvature for the field dependence of the magnetization.

\begin{figure}
\includegraphics[scale=0.9]{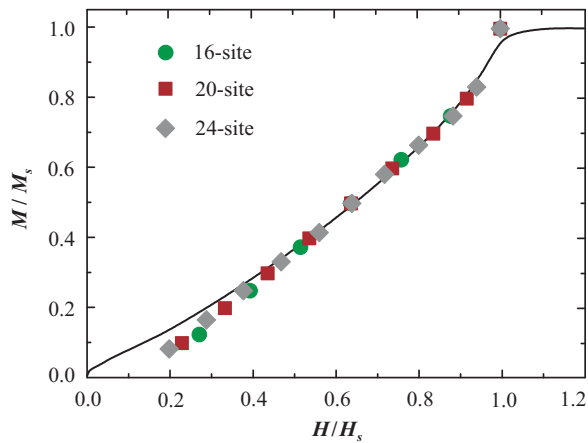}
\caption{\label{comparison}
Experimental magnetization curve for PbZnVO(PO$_4)_2$ along with the FTLM simulations. Circles, squares, and diamonds denote the results for the 16-site, 20-site, and 24-site clusters, respectively.
}
\end{figure}

The agreement between experiment and theory is not only qualitative, but quantitative as well. In Fig.~\ref{comparison}, we plot the experimental data for PZVO along with the simulated magnetization curve. The simulations were done for clusters with 16, 20, and 24 sites using finite-temperature Lanczos method (FTLM).\cite{FTLM} The data points were obtained via the Bonner-Fisher technique.\cite{bonner-fisher} The frustration ratio was taken from the experimental estimate $\bar J_2/\bar J_1=-1.9$. We find perfect agreement between the experimental and simulated curves above $0.4H_s$. At lower fields, the simulated curve lies below the experimental one. This should be related to finite-size effects which are expected to be most pronounced at low fields. Nevertheless, the consistency of the experimental and theoretical results shows that simulations for the relatively small clusters can be an effective tool for studying high-field magnetization of different frustrated spin systems in two dimensions (e.g., the kagom\'e lattice).

In conclusion, we performed high-field magnetization measurements for a number of layered vanadium phosphates representing spin-$\frac12$ FSL compounds. The pulsed-field experiments provide accurate estimates of the saturation fields, while the latter hold important quantitative information on the properties of the FSL systems. The saturation fields of layered vanadium phosphates are consistent with the previous low-field estimates of averaged exchange couplings $\bar J_1$ and $\bar J_2$. This confirms the placement of the compounds on the phase diagram and evidences the strongest frustration in SrZnVO(PO$_4)_2$ and BaCdVO(PO$_4)_2$. Additionally, the values of the saturation fields provide an estimate for the spatial anisotropy of nearest-neighbor couplings. We show that this effect is relatively weak, and the difference between $J_1'$ and $J_1$ does not exceed $1.0-1.5$~K in the whole family of layered vanadium phosphates. The comparison of the magnetization curves for the compounds with different frustration ratios reveals the effect of frustration-induced quantum fluctuations. The frustration leads to the increased bending of the magnetization curves in agreement with the theoretical predictions. The direct comparison of the experimental and simulated curves indicates high accuracy of the simulation techniques and validates further model studies of two-dimensional frustrated spin systems.

Financial support of RFBR (Grant No. 07-03-00890) is acknowledged. We are grateful to Franziska Weickert, Miriam Schmitt, and Oleg Janson for their help in magnetization measurements. A.Ts. acknowledges the hospitality and funding from MPI CPfS. Part of this work has been supported by EuroMagNET under the EU contract RII3-CT-2004-506239.

\end{document}